\documentclass[9pt]{article}
\usepackage{spconf}
\usepackage{amsmath}
\usepackage{graphicx}
\usepackage{tabularx}
\usepackage{lipsum}
\usepackage{booktabs}
\usepackage{multirow}
\usepackage{hhline}
\usepackage{tabularx,booktabs}
\usepackage{float}
\usepackage{epsfig}
\usepackage{latexsym}
\usepackage{psfrag}
\usepackage{pstricks}
\usepackage{dsfont}
\usepackage{cancel}
\usepackage{epstopdf}
\usepackage{bigints}
\usepackage{cite}
\usepackage{url}
\usepackage[normalem]{ulem}
\usepackage{multirow}

\usepackage{amsfonts}
\usepackage{amssymb}

\title{On the transferability of adversarial examples against CNN-based image forensics}
\name{M.Barni, K. Kallas, E. Nowroozi,  B.Tondi
 \thanks{This work has been partially supported by a research sponsored by DARPA and Air Force Research Laboratory (AFRL) under agreement number FA8750-16-2-0173. The U.S. Government is authorised to reproduce and distribute reprints for Governmental purposes notwithstanding any copyright notation thereon. The views and conclusions contained herein are those of the authors and should not be interpreted as necessarily representing the official policies or endorsements, either expressed or implied, of DARPA and Air Force Research Laboratory (AFRL) or the U.S. Government.\newline * The list of authors is provided in alphabetic order.}\vspace{-0.2cm}}
\address{Department of Information Engineering and Mathematics\\University of Siena}

\begin{document}
\ninept

\maketitle
\begin{abstract}
Recent studies have shown that Convolutional Neural Networks (CNN) are relatively easy to attack through the generation of so called adversarial examples. Such vulnerability also affects CNN-based image forensic tools. Research in deep learning has shown that adversarial examples exhibit a certain degree of transferability, i.e., they maintain part of their effectiveness even against CNN models other than the one targeted by the attack. This is a very strong property undermining the usability of CNN's in security-oriented applications. In this paper, we investigate if attack transferability also holds in image forensics applications.
With specific reference to the case of manipulation detection, we analyse the results of several experiments considering different sources of mismatch between the CNN used to build the adversarial examples and the one adopted by the forensic analyst. The analysis ranges from cases in which the mismatch involves only the training dataset, to cases in which the  attacker and the forensic analyst adopt different architectures. The results of our experiments show that, in the majority of the cases, the attacks are not transferable, thus easing the design of proper countermeasures at least when the attacker does not have a perfect knowledge of the target detector.
\end{abstract}

\section{INTRODUCTION}
\label{sec:intro}

Convolutional Neural Networks (CNN) are increasingly used in image forensic applications due to their superior accuracy in detecting a wide number of image manipulations, including multiple JPEG compression \cite{WZ16, barni2017aligned}, median filtering \cite{ChenMFwithCNN}, resizing \cite{BayarCNNUnivTIFS}, contrast manipulation \cite{BCNT18}. Good performance of CNNs have also been reported for image source attribution, i.e., to identify the model of the camera which acquired a certain image \cite{Besta17,FNBC18,BSIcassp18}. Despite the good performance they achieve, the use of CNNs in security-oriented applications, like image forensics, is hindered by the easiness with which adversarial examples can be built \cite{szegedy2013intriguing, Carlini16, Paper16}. As a matter of fact, an attacker who has access to the internal details of the CNN used for a certain image recognition task can easily build an attacked image which is visually indistiguishable from the original one, but is misclassified by the CNN. Such a problem is currently the subject of an intense research activity, yet no satisfactory solution has been found yet (see \cite{AkAj18} for a recent survey on this topic). The problem is worsened by the observation that adversarial attacks are often transferrable from the target network to other networks designed for the same task \cite{PaperTransf16}. This means that even in a Limited Knowledge (LK) scenario, wherein the attacker has only partial information about the to-be-attacked network, he can attack a surrogate network mimicking the target one and the attack will be effective also on the target network with good probability. Such a property opens the way towards very powerful attacks that can be used in real applications wherein the attacker does not have full access to the attacked system \cite{PaperTransf16}.

Following some recent researches, showing that CNN-based image forensics tools are also endangered by the existence of adversarial examples \cite{BestaAdv17, VerdoAdv18}, the goal of this paper is to investigate if and to which extent the transferability of adversarial examples holds in image forensics applications. The answer to this question is of primary importance, since attack transferability would greatly complicate the development of anti-counter-forensics measures. In fact, even denying to the attacker a full access to the forensic tools would not guarantee that the forger can not mislead the forensic analysis. To the best of our knowledge, the only previous work partially addressing this problem is \cite{VerdoAdv18}. In particular, \cite{VerdoAdv18} reports some tests aiming at assessing the transferability of adversarial examples targeting various CNN-based camera model identification systems. According to \cite{VerdoAdv18}, in a camera model identification scenario, attacks are only partially transferable, since the transferred attack succeed in no more than 40\% of the cases (often much less). The analysis in \cite{VerdoAdv18} is a very preliminary one, hence calling for new tests addressing different sources of mismatch between the attacked network and the targeted one, different forensics scenarios, and the impact that attack strength has on the transferability of the attacks. In this paper, we make some steps in this direction. First of all, we consider two forensic tasks boiling down to a binary detection problem. This marks an important difference with respect to \cite{VerdoAdv18}, where the forensic analysis consisted in the classification of the input image into one of several classes rather than in making a binary decision. Specifically we focus on median filtering and image resizing detection. Secondly, we analyse separately the effect of training data mismatch and network architecture mismatch on the transferability of the attacks. We consider two different attack methodologies, namely JSMA \cite{Paper16} and FGSM \cite{goodfellow2014explaining}  and evaluate the transferability of the attacks also in the presence of double-to-integer rounding, which is a necessary step to bring back the attacked image into the integer domain. As we will see, our experiments cast serious doubts on the transferability of adversarial attacks in image forensic applications, thus opening the way to the development of proper countermeasures at least when the attacker does not have a perfect knowledge of the target network.

The rest of this paper is organised as follows. In Sect. \ref{sec.METH}, we describe the methodology used for our experiments, including: i) the description of the algorithms used to generate the adversarial examples; ii) the description of the CNN architectures targeted by the attacks; iii) the description of the experimental campaign, iv) the datasets used for training and testing the CNNs. The results of the experiments are presented in Sect. \ref{sec.EXP}, together with a discussion of our main findings. The paper ends in Sect. \ref{sec.CONC}, where we summarise the lessons we learnt from our experiments and we present a roadmap for future research.

\section{METHODOLOGY}
\label{sec.METH}

In order to evaluate the factors that influence the transferability of adversarial attacks against CNN-based detection of image processing operators, we considered two different kinds of attacks, two detection tasks solved by relying on two different networks, and three sources of mismatch between the network used to create the adversarial attack (hereafter referred to as Source Network - SN) and  the one the attack should be transferred to (hereafter referred to as Target Network - TN).
%
In particular we considered the cases of two different networks trained on the same dataset and the case of a single network trained on different datasets.
With reference to the terminology established in \cite{PaperTransf16}, we refer to the first type of transferability as {\em cross-model transferability} and to the second as {\em cross-training transferability}.
We also considered the case of two different networks trained on different datasets ({\em cross-model-and-training transferability}).
The combination of the above factors resulted in an extensive campaign of experiments whose results will be discussed in Sect. \ref{sec.EXP}.

\subsection{Attacks}

In our experiments, the adversarial examples were built by relying on the Fast Gradient Sign Method (FGSM) algorithm, originally proposed in \cite{goodfellow2014explaining}, and the Jacobian-based Saliency Map Attack (JSMA) \cite{Paper16}.

For the FGSM, we used the refined iterative version (I-FGSM) described in \cite{kurakin2016adversarial}.
In its original implementation, FGSM  obtains an adversarial
perturbation in a computationally efficient way by computing the gradient
of the output with respect to the input image and considering its
sign multiplied by a strength factor $\varepsilon_s$.
The I-FGSM algorithm is a multi-step variant of FGSM; for a given attack strength $\varepsilon_s$, the algorithm is applied iteratively until an adversarial image can be produced (that is, an image which is misclassified by the network), for a maximum number of steps $S$. Several values of $\varepsilon_s$ are considered, i.e. $\varepsilon_s \in E$; the value which minimizes the distortion of the final attacked image with respect to the original one is eventually selected as best strength, for the given maximum number of iterations of the algorithm $S$.

The JSMA algorithm, has been proposed by Papernot et al. in \cite{Paper16} and works as follows:
it consists of a greedy iterative
procedure which relies on forward propagation to compute, at each
iteration, a saliency map, indicating the pixels that contribute most to
the classification. The pixels are then modified based on this map by a relative amount $\theta$, $\theta < 1$ ($\theta$ is relative to the range of the values of the image, the pixel modification being $\theta \times (\max(I) - \min(I))$). A constrain is put on the maximum number of times $T$ the same pixel can be modified. The procedure ends when the attacker succeeds or the pixels are modified by a too large amount (i.e., the number of modifications reaches the maximum prescribed number for all pixels).

In our experiments we used the Foolbox toolbox \cite{rauber2017foolbox} to implement the above attacks.

Both the I-FGSM and the JSMA algorithms produce a real-valued attacked image. While in some cases we can assume that the attacked image is used as is, in most applications image pixels must be mapped back into the integer domain before being fed to the CNN. This may result in a loss of effectiveness of the attack, since some of the subtle changes introduced by the attack are deleted when pixels are rounded (or truncated) to integer values. 
%

\subsection{Datasets}

In order to evaluate the transferability of the attacks when the SN and the TN are trained on different datasets, we considered the RAISE (R)\cite{RAISE8K} dataset and the VISION (V) dataset \cite{Shull17}.

For our experiments, about 2000 uncompressed, camera-native, images (.tiff) were taken from the
RAISE dataset, with size $4288\times2848$. These images are camera-native images coming from three different cameras.
The same number of images were taken from the VISION dataset. This dataset consists of
native images acquired by smartphones/tablets belonging to several brands. To get similar resolution images for the two datasets, we only selected the devices for which the resolution was not very different from that of the images from RAISE. Specifically, the sizes of the images we considered ranges from a minimum of $2336\times4160$ up to $3480\times4640$.
The images from the VISION dataset are in JPEG format. In order to reduce the possible impact of compression artefact, we selected images only from the high-quality devices, for which the JPEG Quality Factor is larger than 97.

The images from both $R$ and $V$ datasets were split into training (and validation) set and test set, and then processed to produce the images for the manipulated class, namely, median and resizing.
For all our tests we considered one-channel images, then all the images from $R$ and $V$ were converted to gray-scale.

%
%
%

\subsection{Networks}
In our experiments, we considered two different detection tasks, namely detection of image resizing (downsampling, by a 0.8 factor) and detection of median filtering (by a 5 $\times$ 5 window). To cope with them, we built several networks generally indicated as $N_{\text{ar}}^{\text{tr}}(\text{task})$, where "ar" indicates the architecture of the network, "tr" $\in$  \{R, V\} the dataset used for training and "task" $\in$ \{med, res\} the detection task ("med" indicating median filtering and "res" resizing).

With regard to the architectures, we considered the network in \cite{bayar2016deep} (recently extended in \cite{BayarCNNUnivTIFS}), hereafter referred to as BSnet ("ar" = BS), and the one in \cite{BCNT18}, hereafter denoted as GCnet ("ar" = GC).
%
%
%
BSnet, originally proposed 
for image manipulation detection and classification, consists of 3 convolutional layers, 3 max-pooling layers and 3 fully-connected layers.
Residual-based features are extracted by constraining the filters of the first layer (with $5 \times 5$ receptive field),  by enforcing
a high-pass nature of the filters (see \cite{BayarCNNUnivTIFS} for more details). For the second and third convolutional layers the filter size is set to $7\times 7$ and $5 \times 5$ respectively, and the stride is set to 2.
For the max-pooling, a kernel size $3 \times 3$ is used with stride 2.

%
%
%
%
%

GCnet was originally proposed to detect generic contrast adjustment operators. With respect to BSnet, GCnet is significantly deeper, consisting of 9 convolutional layers. The network has only 2 max-pooling layers and one fully-connected layer.
A kernel size of $3 \times 3$ and stride 1 was used for all the convolutional
layers. Max-pooling is applied with kernel size $2 \times 2$ and stride 2.
The number of parameters is then reduced by halving the number
of feature maps in the final convolutional layer, and considering just one fully-connected layer.


%

In summary, we built 6 networks, indexed as follows: $N_{\text{BS}}^{\text{tr}}(\text{task})$, "tr" $\in$ \{R, V\}, "task" $\in$ \{med, res\}, and $N_{\text{GC}}^{\text{R}}(\text{task})$ with  "task" $\in$ \{med, res\}.

\subsection{Experiments}
\label{sec.ExpSetup}

The experimental campaign was designed in such a way to highlight attack transferability in a wide variety of settings. Experiments have been split into three categories according to the type of mismatch between the SN and the TN. We started studying cross-model transferability, according to which SN and TN share the same architecture, but are trained on different datasets. Then we passed to analyse cross model transferability, in which different network architectures are trained on the same dataset.  Eventually, we passed to cross-model-and-training transferability according to which the SN and the TN share neither the architecture nor the training data. All the tests have been repeated for both resizing and median filtering detection. For sake of simplicity we did not consider all possible combinations, however, as it will be evident from the subsequent section, the amount of experiments we carried out is sufficient to draw a number of significant conclusions. 
%
In particular, the experiments for the cross-training transferability are carried out by considering only BSnet as the SN, trained on both $R$ (i.e., $SN = N_{\text{BS}}^{\text{R}}$) and $V$ (i.e., $SN = N_{\text{BS}}^{\text{R}}$). For the experiments on the model-training transferability, BSnet is taken as SN and GCnet as TN, both trained on $R$ (i.e.,  $SN = N_{\text{BS}}^{\text{R}}$ and $TN = N_{\text{GC}}^{\text{R}}$). Finally, for the  cross-model-and-training case, we set $SN = N_{\text{BS}}^{\text{V}}$  and $TN = N_{\text{GC}}^{\text{R}}$.

With regard to the attacks, for the I-FGSM attack, the number of steps $S$ is fixed to 10 (default). The best strength is searched in the range  $E  = [0: \varepsilon_s :0.1]$, where $\varepsilon_s$ is the search step size, which also corresponds to the minimum strength considered. Setting a larger $\varepsilon_s$ generally corresponds to consider a stronger attack. In our experiments, we considered $\varepsilon_s = 0.001$ and  $0.01$, which satisfy the condition that the average PSNR remains above 40 dB.

For the JSMA, $T$ is set to 7. The algorithm is applied with the default maximum number of iterations 2000. The relative modification per pixel $\theta$ is set to 0.01 and 0.1, the second case corresponding to a stronger attack. We did not consider $\theta$ values larger than 0.1, since above this value the maximum pixel distortion introduced by the attack starts becoming too large ($> 70$).

Eventually, we repeated all the experiments by rounding the output of the attack to integer values.

\section{RESULTS AND DISCUSSION}
\label{sec.EXP}

In this section we discuss the results of the experiments we have carried out, by considering separately the cases of cross-training, cross-model and cross-model-and-training transferability. For sake of brevity, we will focus on the floating point version of the attacks, being this case more favorable to the attacker, and we will briefly touch upon the integer-valued case at the end of the section.

\subsection{Training}

To build our models $N_{\text{BS}}^{\text{R}}$ and $N_{\text{BS}}^{\text{V}}$ (for both detection tasks),
we considered 200.000 patches for training (and validation) and 10000 for testing, per class. The input patch size is set to $128 \times 128$ in all the cases.
In order to use all the images in the datasets $R$ and $V$, a maximum number of 100 patches is selected (randomly) for each image.
A number of 30 training epochs was considered (as in \cite{bayar2016deep}).
For the deeper models $N_{\text{GC}}^{\text{R}}$ for both the "med" and "res" task, we used $10^6$ patches for training, $10^5$ for validation, and $5*10^4$ for testing. To reach these numbers, all the image patches were selected from all the images. By following \cite{BCNT18}, the number of training epochs is set to 3.
For training both BSnet and GCnet, the Adam solver
is used with learning rate $10^{-4}$ and momentum $0.99$. The
batch size for training and validation is set to 32 images, the test batch size to 100.
%
The accuracies achieved by the BSnet in absence of attacks in the various cases are: $98.1\%$ for $N_{\text{BS}}^{\text{R}}(\text{med})$, $99.5 \%$ for $N_{\text{BS}}^{\text{V}}(\text{med})$, $97.5\%$ for $N_{\text{BS}}^{\text{R}}(\text{res})$, $96.6 \%$ for $N_{\text{BS}}^{\text{V}}(\text{res})$.
With regard to GCnet, it got the following accuracies: $98.4 \%$ for $N_{\text{GC}}^{\text{R}}(\text{med})$ and $98.5 \%$ for $N_{\text{GC}}^{\text{R}}(\text{res})$.

In the next section, we discuss the performance of the models in the presence of attacks, in the matched and mismatched cases. 
In all the cases, the performance are assessed on 500 attacked images.

\subsection{Cross-training transferability}

As detailed in Sect. \ref{sec.ExpSetup}, these experiments were carried out by considering only the BS architecture.
The results we got are reported in Table
\ref{tab:crossTR}. For all the cases, the PSNR, $L_1$ distortion and maximum absolute distortion  are reported, averaged on all the images successfully attacked in the matched case.
As we can see, the attacks are generally non-transferable and the images attacked using SN are not able to deceive the TN. More specifically, with the FGSM attack, the adversarial examples can be transferred in a significant number of  cases only when the larger strength is considered ($\varepsilon_s = 0.01$)
and the SN corresponds to $N_{\text{BS}}^{\text{R}}(\text{res})$ and $N_{\text{BS}}^{\text{R}}(\text{med})$ (attack success rate $0.6923$ and $0.8452$ respectively) and to $ N_{\text{BS}}^{\text{V}}(\text{res})$ (attack success rate $0.9415$). For the JSMA case, the attack can be transferred only when SN is $ N_{\text{BS}}^{\text{R}}(\text{res})$ and strong attack with $\theta = 0.1$ is considered, with success rate 0.7821. Furthermore, we observe that the JSMA is never transferable when the VISION dataset is used to train the SN. 

According to our tests, attacks obtained by JSMA are less transferable than those produced by FGSM. A possible motivation can be the following: since very few pixels are modified by JSMA (those which the network is more sensitive to), it tends to overfit more the attacked model. It is also interesting to observe that, for a given detection task, the transferability is not symmetric with respect to the datasets used for training. This suggests that, in forensic applications, the features learned by the network  may also be affected in some way and up to some extent by the underlying dataset. This point deserves further investigation as a future work.

\begin{table*}[!htbp]
\caption{Experimental results for Cross Training case. Transferable attacks are highlighted in bold.}
\label{tab:crossTR}
\resizebox{\textwidth}{!}{\begin{tabular}{|c|c|c|c|c|c|c|c|c|}
\hline
\multicolumn{9}{|c|}{\textbf{CROSS TRAINING}}   \\ \hline

SN &TN  & accuracy   & attack type &avg. PSNR  &avg. L1 dist  &avg. max. dist  &attack success rate on SN  &attack success rate on TN  \\ \hline

$N_{\text{BS}}^{\text{R}}(\text{res})$ & $N_{\text{BS}}^{\text{V}}(\text{res})$  &  SN= 97.60\%, TN= 96.00\%  & I-FGSM,  $\varepsilon_s= 0.01$ & 40.02 & 2.53 & 2.55 & \textbf{1.0000} & \textbf{0.6923} \\ \hline

$N_{\text{BS}}^{\text{R}}(\text{res})$ & $N_{\text{BS}}^{\text{V}}(\text{res})$  &  SN=97.60\%, TN= 96.00\%  & I-FGSM,  $\varepsilon_s=0.001$ & 58.46 & 0.26 & 0.27 & 1.0000 & 0.0491 \\ \hline

$N_{\text{BS}}^{\text{R}}(\text{res})$ & $N_{\text{BS}}^{\text{V}}(\text{res})$  &  SN= 97.60\%, TN= 96.00\%  & JSMA,  $\theta=0.1$ & 46.04 & 0.07 & 58.32 & \textbf{1.0000} & \textbf{0.7821} \\ \hline

$N_{\text{BS}}^{\text{R}}(\text{res})$ & $N_{\text{BS}}^{\text{V}}(\text{res})$  &  SN= 97.60\%, TN= 96.00\%  & JSMA,  $\theta=0.01$ & 54.99 & 0.04 & 15.09 & 0.99 & 0.11 \\ \hline

$N_{\text{BS}}^{\text{V}}(\text{res})$ & $N_{\text{BS}}^{\text{R}}(\text{res})$  &  SN= 97.80\%, TN= 99.60\%  & I-FGSM,  $\varepsilon_s=0.01$ & 40.03 & 2.53 & 2.55 & 1.0000 & 0.0021 \\ \hline

$N_{\text{BS}}^{\text{V}}(\text{res})$ & $N_{\text{BS}}^{\text{R}}(\text{res})$  &  SN= 97.80\%, TN= 99.60\%  & I-FGSM,  $\varepsilon_s=0.001$ & 59.64 & 0.26 & 0.27 & 1.0000 & 0.0000 \\ \hline

$N_{\text{BS}}^{\text{V}}(\text{res})$ & $N_{\text{BS}}^{\text{R}}(\text{res})$  &  SN= 97.80\%, TN= 99.60\%  & JSMA,  $\theta=0.1$ & 50.55 & 0.01 & 69.42 & 0.99 & 0.00 \\ \hline

$N_{\text{BS}}^{\text{V}}(\text{res})$ & $N_{\text{BS}}^{\text{R}}(\text{res})$  &  SN= 97.80\%, TN= 99.60\%  & JSMA,  $\theta=0.01$ & 57.78 & 0.01 & 17.06 & 0.98 & 0.0000 \\ \hline \hline

$N_{\text{BS}}^{\text{R}}(\text{med})$ & $N_{\text{BS}}^{\text{V}}(\text{med})$  &  SN= 98.20\%, TN= 100\%  & I-FGSM,  $\varepsilon_s=0.01$ & 40.03 & 2.53 & 2.55 & \textbf{1.0000} & \textbf{0.8452} \\ \hline

$N_{\text{BS}}^{\text{R}}(\text{med})$ & $N_{\text{BS}}^{\text{V}}(\text{med})$  &  SN= 98.20\%, TN= 100\%  & I-FGSM,  $\varepsilon_s= 0.001$ & 59.67 & 0.26 & 0.27 & 1.00 & 0.04 \\ \hline

$N_{\text{BS}}^{\text{R}}(\text{med})$ & $N_{\text{BS}}^{\text{V}}(\text{med})$  &  SN= 98.20\%, TN= 100\%  & JSMA,  $\theta=0.1$ & 49.64 & 0.03 & 38.11 & 1.0000 & 0.0122 \\ \hline

$N_{\text{BS}}^{\text{R}}(\text{med})$ & $N_{\text{BS}}^{\text{V}}(\text{med})$  &  SN= 98.20\%, TN= 100\%  & JSMA,  $\theta=0.01$ & 58.47 & 0.02 & 14.05 & 0.98 & 0.0020 \\ \hline

$N_{\text{BS}}^{\text{V}}(\text{med})$ & $N_{\text{BS}}^{\text{R}}(\text{med})$  &  SN= 100\%, TN= 99.20\%  & I-FGSM,  $\varepsilon_s=0.01$ & 40.04 & 2.53 & 2.55 & \textbf{1.0000} & \textbf{0.9415} \\ \hline

$N_{\text{BS}}^{\text{V}}(\text{med})$ & $N_{\text{BS}}^{\text{R}}(\text{med})$  &  SN= 100\%, TN= 99.20\%  & I-FGSM,  $\varepsilon_s=0.001$ & 59.94 & 0.25 & 0.25 & 1.0000 & 0.07 \\ \hline

$N_{\text{BS}}^{\text{V}}(\text{med})$ & $N_{\text{BS}}^{\text{R}}(\text{med})$  &  SN= 100\%, TN= 99.20\%  & JSMA,  $\theta=0.1$ & 49.55 & 0.03 & 32.09 & 1.0000 & 0.0101 \\ \hline

$N_{\text{BS}}^{\text{V}}(\text{med})$ & $N_{\text{BS}}^{\text{R}}(\text{med})$  &  SN= 100\%, TN= 99.20\%  & JSMA,  $\theta=0.01$ & 58.13 & 0.01 & 14.08 & 0.9879 & 0.0081 \\ \hline
\end{tabular}}
\end{table*}
\begin{table*}[!h]
\caption{Experimental results for Cross Model case.  Transferable attacks are highlighted in bold.}
\label{tab:crossMD}
\resizebox{\textwidth}{!}{\begin{tabular}{|c|c|c|c|c|c|c|c|c|}
\hline
\multicolumn{9}{|c|}{\textbf{CROSS MODEL}}   \\ \hline
SN &TN  &Accuracy w/o attack  &attack  &avg. PSNR  &avg. L1 dist  &avg. max. dist  &attack success rate on SN  &attack success rate on TN  \\ \hline

$N_{\text{BS}}^{\text{R}}(\text{res})$ & $N_{\text{GC}}^{\text{R}}(\text{res})$  &  SN= 97.60\%, TN= 98.20\%  & I-FGSM,  $\varepsilon_s=0.01$ & 40.02 & 2.53 & 2.55 & 1.0000 & 0.0020 \\ \hline

$N_{\text{BS}}^{\text{R}}(\text{res})$ & $N_{\text{GC}}^{\text{R}}(\text{res})$  &  SN= 97.60\%, TN= 98.20\%  & I-FGSM,  $\varepsilon_s=0.001$ & 58.48 & 0.31 & 0.33 & 1.0000 & 0.0020 \\ \hline

$N_{\text{BS}}^{\text{R}}(\text{res})$ & $N_{\text{GC}}^{\text{R}}(\text{res})$  &  SN= 97.60\%, TN= 98.20\%  & JSMA,  $\theta=0.1$ & 46.09 & 0.07 & 57.88 & 1.0000 & 0.0164 \\ \hline

$N_{\text{BS}}^{\text{R}}(\text{res})$ & $N_{\text{GC}}^{\text{R}}(\text{res})$  &  SN= 97.60\%, TN= 98.20\%  & JSMA,  $\theta=0.01$ & 54.98 & 0.04 & 15.14 & 0.9918 & 0.0061 \\ \hline \hline

$N_{\text{BS}}^{\text{R}}(\text{med})$ & $N_{\text{GC}}^{\text{R}}(\text{med})$  &  SN= 98.20\%, TN= 100\%  & I-FGSM,  $\varepsilon_s=0.01$ & 40.03 & 2.53 & 2.55 & \textbf{1.0000} & \textbf{0.8248} \\ \hline

$N_{\text{BS}}^{\text{R}}(\text{med})$ & $N_{\text{GC}}^{\text{R}}(\text{med})$  &  SN= 98.20\%, TN= 100\%  & I-FGSM,  $\varepsilon_s= 0.001$ & 59.67 & 0.26 & 0.27 & 1.0000 & 0.1813 \\ \hline

$N_{\text{BS}}^{\text{R}}(\text{med})$ & $N_{\text{GC}}^{\text{R}}(\text{med})$  &  SN= 98.20\%, TN= 100\%  & JSMA,  $\theta=0.1$ & 49.64 & 0.03 & 38.11 & 1.0000 & 0.0102 \\ \hline

$N_{\text{BS}}^{\text{R}}(\text{med})$ & $N_{\text{GC}}^{\text{R}}(\text{med})$  &  SN= 98.20\%, TN= 100\%  & JSMA,  $\theta=0.01$ & 58.47 & 0.02 & 14.05 & 0.9837 & 0.0163 \\ \hline
\end{tabular}}
\end{table*}
\begin{table*}[!h]
\caption{Experimental results for Cross Training and Model case.  Transferable attacks are highlighted in bold.}
\label{tab:crossEVERY}
\resizebox{\textwidth}{!}{\begin{tabular}{|c|c|c|c|c|c|c|c|c|}
\hline
\multicolumn{9}{|c|}{\textbf{CROSS-TRAINING-AND-MODEL}}   \\ \hline
SN &TN  &Accuracy w/o attack  &attack  &avg. PSNR  &avg. L1 dist  &avg. max. dist  &attack success rate on SN  &attack success rate on TN  \\ \hline

$N_{\text{BS}}^{\text{V}}(\text{res})$ & $N_{\text{GC}}^{\text{R}}(\text{res})$ & SN= 99.20\%, TN= 99.60\% & I-FGSM,  $\varepsilon_s=0.01$ & 40.03 & 2.53 & 2.55 & 1.0000 & 0.0040 \\ \hline

$N_{\text{BS}}^{\text{V}}(\text{res})$ & $N_{\text{GC}}^{\text{R}}(\text{res})$ & SN= 99.20\%, TN= 99.60\% & I-FGSM,  $\varepsilon_s=0.001$ & 59.57 & 0.27 & 0.27 & 1.0000 & 0.0020 \\ \hline

$N_{\text{BS}}^{\text{V}}(\text{res})$ & $N_{\text{GC}}^{\text{R}}(\text{res})$ & SN= 99.20\%, TN= 99.60\% & JSMA,  $\theta=0.1$ & 50.20 & 0.02 & 70.87 & 1.0000 & 0.0000 \\ \hline

$N_{\text{BS}}^{\text{V}}(\text{res})$ & $N_{\text{GC}}^{\text{R}}(\text{res})$ & SN= 99.20\%, TN= 99.60\% & JSMA,  $\theta=0.01$ & 57.40 & 0.01 & 17.16 & 0.9919 & 0.0000 \\ \hline \hline

$N_{\text{BS}}^{\text{V}}(\text{med})$ & $N_{\text{GC}}^{\text{R}}(\text{med})$ & SN= 100\%, TN= 100\% & I-FGSM,  $\varepsilon_s=0.01$ & 40.04 & 2.53 & 2.55 & \textbf{1.0000} & \textbf{0.7960} \\ \hline

$N_{\text{BS}}^{\text{V}}(\text{med})$ & $N_{\text{GC}}^{\text{R}}(\text{med})$ & SN= 100\%, TN= 100\% & I-FGSM,  $\varepsilon_s= 0.001$ & 59.91 & 0.25 & 0.26 & 1.0000 & 0.0080 \\ \hline

$N_{\text{BS}}^{\text{V}}(\text{med})$ & $N_{\text{GC}}^{\text{R}}(\text{med})$ & SN= 100\%, TN= 100\% & JSMA,  $\theta=0.1$ & 49.56 & 0.03 & 31.83 & 1.0000 & 0.0080 \\ \hline

$N_{\text{BS}}^{\text{V}}(\text{med})$ & $N_{\text{GC}}^{\text{R}}(\text{med})$ & SN= 100\%, TN= 100\% & JSMA,  $\theta=0.01$ & 58.06 & 0.01 & 14.18 & 0.9900 & 0.0120 \\ \hline
\end{tabular}}
\end{table*}

\vspace{-0.1cm}
\subsection{Cross-model transferability}

In this case, the experiments where carried out by considering only the $R$ dataset and using the BS architecture for the SN. The results we have got are reported in Table
\ref{tab:crossMD}.
The experiments show the lack of transferability with respect to a mismatch in the network model. The only exception is  for the "med" case, in which case the stronger attack (with $\varepsilon_s = 0.01$) is transferable 82.5\% of the times. However, it is worth stressing that, when the FGSM is applied with such a strength, although the PSNR is not very low (40.03 dB), the average $L_1$ distortion is around 2.5 (a similar value is attained by the maximum absolute distortion). With such values of $L_1$, the visual quality of the FGSM attacked images is impaired and peculiar visual artifacts appears, especially in relatively uniform image patches.

The fact that the lack of transferability is even stronger in the "res" case than in the "med" case can be probably justified by the ease of the median filtering detection task (even because the median filtering is performed with a rather large window size), compared to the resize. Therefore, we might expect that in the case of "med" similar peculiar features are learned by the shallow and deeper network, hence facilitating the transferability of the attacks.

\subsection{Cross-model-and-training transferability}

In this case, the experiments were carried out by considering the BS architecture trained on the $V$ dataset as the SN, and the GC architecture trained on the $R$ dataset as the TN. Similar results can be obtained by combining architecture and dataset in the other way round. The results we have got are reported in Table \ref{tab:crossEVERY}. Quite expectedly, the table shows that the transferability of the attacks in this case decreases further and the attack success rate is below 0.01 in all the cases but for the case of FGSM with $\varepsilon_s = 0.01$, for which a success rate of 0.796 can still be achieved.

Lastly, we repeated all the experiments by rounding the pixel values of the attacked images to integers. According to the results we have got, integer rounding does not have a big impact on the transferability of the attacks. Rather it influences the effectiveness of the attack on the SN itself, as already reported in several studies including \cite{VerdoAdv18} and \cite{Tondi}.

\section{CONCLUDING REMARKS}
\label{sec.CONC}
\vspace{-0.1cm}
We investigated the transferability of adversarial examples in an image forensics scenario. By focusing on two manipulation detection tasks, we run tests by considering two well known attack methodologies and several sources of mismatch. Our tests show that adversarial examples are generally non-transferable, in contrast to what happens in typical patter recognition applications.
This states an important result, since the lack of transferability can be exploited by the forensic analyst to make the attack more difficult. For instance, a LK scenario can be enforced in some way to combat adversarial examples, as done with the approaches based on standard ML. Even if our results clearly show that adversarial examples can not be easily transferred from one network to another, further tests are needed before we can draw some final conclusions. First of all, more detection tasks should be considered, together with different sources of mismatch between the SN and the TN. As an example, we may wonder if a mismatch in the training procedure is enough to prevent transferability.  Also, the reason why image-forensic networks are less prone to attack transfers should be understood.  
On the attacker's hand, new research is needed to understand how the transferability could be improved by increasing the attack strength, especially when the adversarial examples must be mapped back in the integer domain.

\bibliographystyle{IEEEbib}
\bibliography{ICASSP19}

\end{document}